# Is it possible for a superluminal signal velocity without violating causality?

Li-Gang Wang [1, 2] [*]

[1] *Department of Physics, Zhejiang University, Hangzhou, 310027, China*

[2] *Department of Physics, The Chinese University of Hong Kong, Shatin, N. T., Hong Kong*

We present a model for a vacuum-like effective medium composed of the absorbing and gain media under the special designed parameters. Within the linear response theory, we prove that any pulse signal (with or without a discontinuity) through such a kind of vacuum-like effective media is always equal to the light speed in vacuum ($c$) without any distortion. As well known that the group velocity in anomalous or normal dispersive media may be smaller or larger than $c$, or even become negative, but the discontinuous point always propagates at the velocity $c$. Therefore we present some discussions on different definitions of the signals, based on the light pulses with a well-defined shape or with a sudden change, for trying to understand two possibilities for the signal velocity without violating the causality.

---

[*] E-mail: sxwlg@yahoo.com.cn

It is well known that the group velocity of a light pulse can be greatly changed by tailoring the dispersive properties of various materials [1, 2, 3, 4]. It has been observed in experiments that the group velocity can be faster than the light speed $c$ and even become negative in anomalous dispersive media ("fast-light" media). Therefore, the group velocity for describing pulse propagations is *at least* measurable and physical meaningful. Another fact is that extreme slow-light phenomena with nearly stopping group velocity have also been observed in "slow-light" media such as an ultracold atomic gas [5] and solid-state materials [6, 7]. For ultraslow group-velocity propagations, there is almost no controversy on signal velocity, while for fast-light propagations, the controversy remains and unsolves [1, 2, 3, 4, 8, 9, 10, 11, 12, 13, 14, 15]. The key problem is "whether could signal velocity be faster than light speed $c$ or not without violating causality?" It is very superising that the core of this controversy is *not the transmission channel itself* (i. e., anomalous or normal dispersive media) *but the definition (or nature) of the true signal*. It is *the definition of the signal not the transmission channel* that determines the signal velocity.

Nowadays, there are two totally different definitions on a signal: one is based on a smooth pulse with a definite bandwidth, and another is defined by a nonanalytical point involved with an infinite bandwidth. These two inconsistent definitions lead to different viewpoints: (1) a signal encoded on a smooth pulse propagates at the group velocity $v_g$, *at least* in the situations of normal dispersion media without doubt [3, 4, 12, 13, 14, 15], and (2) a signal encoded onto the nonanalytical point travels at $c$ [1, 8, 16, 17, 18, 19]. For the first viewpoint, because $v_g$ is related with the group index $n_g$ that is frequency dependent, so the signal velocity ($v_s = v_g$) actually depends on the dispersive properties of the transmission channel, and *different media may have different signal velocities*. This viewpoint was generally acceptable and applicable in many conventional signal communication systems. For the second viewpoint, the signal encoded on a nonanlytical point does always propagate at $c$ in any medium. In particular, Stenner et al. [8, 16] experimentally estimated the speed of a *rapid* unpredictable change in the pulse amplitude be the light speed $c$ in both the fast-light and slow-light media. Their experiments astonishedly tells people that *the signla velocity is independent of transmission channel*. In reality, we find that this viewpoint is not the case in the communication systems including the digital-signal systems, in which the signal velocity always depends on transmission channels (media). We have questions: if a signal is encoded by an discontinuous point, then *how to* use this kind of signal in real communication systems? where signal velocity is independed of any materials, and if the first viewpoint is correct, *how to* make it possible be compatible with the causality?

The motivation of this paper is to make a clear assessment of the above two signal velocities. Our proposal presented in this paper may or may not turn out to be correct (or interesting), but it is at least unambiguous, and might be tested in certain ways to be valuable for the controversy. First we present a simple proposal for obtaining a vacuum-like effective medium, consisted of an absorbing medium and a gain medium. It can be proved that any type of signals through such a kind of vacuum-like effective media is always equal to $c$ without any distortion. Then we give out some detailed discussions on the two possibilities of signal velocities in dispersive media.

We first consider the propagation of a light pulse with a carrier frequency $\omega_0$ passing through an absorbing medium and a gain medium, as shown in Fig. 1. Both the absorbing and gain media have the same lengths, $L/2$. The susceptibilities of the absorbing and gain dispersive media are the form of double Lorentz oscillators

$$\chi_{a,g}(\omega) = \frac{M_{a,g}}{\omega - \omega_0 - \Delta_{a,g} + i\gamma_{a,g}} + \frac{M_{a,g}}{\omega - \omega_0 + \Delta_{a,g} + i\gamma_{a,g}}, \qquad (1)$$

under different parameters: $M_a < 0$ for the absorbing medium and $M_g > 0$ for the gain medium. The subscripts "a" and "g" denote the absorbing and gain media, respectively, $\Delta_{a,g}$ denote the detuning of the resonant frequencies of the oscillators to the carrier frequency, and $\gamma_{a,g}$ denote the decay rates of the oscillators. Equation (1) could be realized in a three-level atomic system with two closely placed Raman aborsbing or gain process [20, 21, 22]. From Eq. (1), the complex refractive indices of both the absorbing and gain media can be given by the relation: $n_{a,g}(\omega) = \sqrt{1 + \chi_{a,g}(\omega)}$.

Due to $n_{a,g}(\omega) \approx 1$ in atomic media, the reflection fields from the interfaces between different media could be negligible. According to the linear response theory, the output pulse (at position $z = L/2$) after the absorbing medium can be readily given by [23]

$$E_{out,1}(z = L/2, t) = \frac{1}{2\pi} \int_{-\infty}^{\infty} E_{in}(0, \omega) \exp\left[i\frac{\omega n_a(\omega) L/2}{c} - i\omega t\right] d\omega, \quad (2)$$

where $E_{in}(0, \omega)$ is the spectrum of the incident pulse $E_{in}(0, t)$ at $z = 0$, and the pulse spectrum and its amplitude have the relation: $E_{in}(0, \omega) = \int_{-\infty}^{\infty} E_{in}(0, t) \exp[i\omega t] d\omega$. The output pulse (at $z = L$) after the gain medium could be given by

$$E_{out,2}(z = L, t) = \frac{1}{2\pi} \int_{-\infty}^{\infty} E_{in}(0, \omega) \exp\left[i\frac{\omega \left[n_a(\omega) + n_g(\omega)\right]}{2} L - i\omega t\right] d\omega, \quad (3)$$

Therefore, from Eq. (3), the absorbing medium and the gain medium may be equivalent to an effective medium with the averaged refractive index $\tilde{n}(\omega) = [n_a(\omega) + n_g(\omega)]/2$ and total thickness $L$. Such equivalent medium has been used for designing an effective medium with particular dispersive properties [24]. In our situation, in the atomic gaseous media, due to $\chi_{a,g}(\omega) \ll 1$, we have $n_{a,g}(\omega) \cong 1 + \frac{1}{2}\chi_{a,g}(\omega)$. Thus the averaged refractive index is given by $\tilde{n}(\omega) = 1 + \frac{1}{2}[\chi_a(\omega) + \chi_g(\omega)]$. When we take the parameters as follows: $\Delta_a = \Delta_g$, $\gamma_a = \gamma_g$, $M_a = -M_g$, we have $\chi_a(\omega) = -\chi_g(\omega)$. Therefore the averaged refractive index of the combined media is given by $\tilde{n}(\omega) = 1$. It indicates that it is possible to have a *vacuum-like* effective medium if we combine the absorbing and gain media under special parameters. When a light pulse passes through such vacuum-like effective media, its shape and amplitude will keep the same as it passes through a same-distance vacuum. From Eq. (3), using the above approximation and condition, the output pulse could be always expressed as

$$E_{out,2}(z = L, t) = E_{in}(0, t - L/c). \quad (4)$$

From Eq. (4), it is proved that the output pulse after through a pair of the absorptive and gain media can be *equivalent to* the situation where it passes through a vacuum with the same distance $L$. Therefore, we have confident in that the pulse signal velocity, for both two definitions described in the above, after passing through such a pair of absorptive and gain media, is ***absolutely*** equal to $c$, which is independent of the pulse shape (whatever the pulse is smooth or has a sudden change).

Then the question arises, "what speed does a practical pulse signal inside the normal or anomalous dispersive media propagate at?" In our following examples, the pulse first passes through a normal dispersive absorbing medium and then passes through an anomalous dispersive gain medium. Then the question becomes that, "what speed does a pulse signal inside the first normal medium? And what speed does a pulse signal inside the second anomalous medium?" Remember that the total speed from z=0 to z=L is equivalent to c.

In order to answer the above questions, we turn to present the strict numerical simulation on the pulse evolution based on Eqs. (2) and (3). Figure 2 shows the typical evolution of a smooth pulse passing through both the absorbing and gain media. It shows that the evolution of the pulse profile from $z=0$ to $z=30$ cm is subliminal at the group velocity $v_g \approx c/311.7$, and the peak time shifts subliminally from the points A to B with a positive delay $\tau_d \approx 0.3117$ μ s, due to the normal dispersion of the absorbing medium. It is also noticed that the relative intensity gradually decays due to the absorption effect. However, from Figs. 2(b) to 2(c), because the pulse spectrum is within the anomalous dispersion of the gain medium, the pulse propagation from $z=30$ cm to $z=60$ cm becomes superluminal with $v_g \approx -c/309.7$. Therefore the peak time shifts from the points B to C with a negative delay $\tau_d \approx -0.3097$ μ s, at the same time, the pulse intensity is gradually amplified to be the same as that of the pulse through the vacuum [Note that the solid and dashed lines are overlapped in Fig. 2(c)]. From Figs. 2(a) to 2(c), it is clearly shown that the final output pulse, see Fig. 2(c), after passing through both the slow-light and fast-light media, is restored to be the same with the original pulse as if it passes through a same-distance vacuum. Therefore we conclude that the signal velocity passing through the combined media of Fig. 1 is ***absolutely*** equal to the speed of light in vacuum, $c$, with no distortion.

Figure 3 show another example for the evolution of a light pulse, with a sudden change $Q$, passing through both the absorbing and gain media. In this case, the absorbing media is an anomalous dispersive medium, and the gain medium becomes a normal dispersive medium. In Fig. 3, the evolution of the pulse peak in the anomalous absorbing medium is superluminal and its peak time moves with the fast-light group velocity ($v_g \approx -c/170.8$) from the points A to B with the attenuated intensity, and from $z = 30$ cm to $z = 60$ cm, the pulse evolution becomes subliminal and the pulse intensity is gradually amplified, and at the same time the peak time moves from the points B to C with the slow-light group velocity $v_g \approx c/172.8$. Similar to the case in Fig. 2, the final output pulse in Fig. 3(c) has the same properties with that through the same-distance vacuum. From Fig. 3 (a) to (c), it should be also noticed that, the sudden change $Q$ is always propagating at a speed of light $c$, whatever the medium is normal or anomalous. This result is in a agreement with the popular arguments in scientific literatures [1, 2, 8, 16, 18, 19], where one believes that a front or a sudden change in a pulse is a "true" signal and the "true" signal velocity is limited to $c$. In Fig. 3(b), it is seen that when the pulse passes through the first medium, the shape of the pulse is strongly changed/distorted near the point of the sudden change $Q$. As the pulse continues to pass through the second gain normal medium, it is clear that its shape including the sudden jump and its amplitude can be recovered as if it passes through a vacuum with the same distance.

Now we back to discuss the possibilities of the above two signal velocities. We know, in any real communication system, it is well believed that a pulse signal inside a normal dispersive media has its signal velocity equal to the group velocity. This argument is widely applied and accepted in modern communication systems [12]. For example, in usual fiber communication systems, one has known that a pulse signal has a signal velocity (equal to $c/n_g$) much smaller than the light speed $c$ (even smaller than the phase velocity $c/n$), and this viewpoint is popular for various kinds of pulse signals with well-defined shapes in the modern commercial communication systems including digital-communication systems, here $n_g$ is the group index and $n$ refers to the host refractive index of a medium. Clearly, if the statement that the pulse signal velocity from z=0 to z=30cm in Fig. 2 (or from z=30cm to z=60cm in Fig. 3) is less than $c$ is assumed to be correct, then it is necessary to require that the signal velocity from z=30cm to z=60cm in Fig. 2 (or from z=0 to z=30cm in Fig. 3) should be presumably faster than $c$, in order to have the total signal velocity equal $c$, as shown in Figs. 2 and 3. Otherwise it is impossible that the final output signal can be recovered with the same delay as in the case through the vacuum (see Figs. 2 and 3).

In Refs. [25, 26, 27, 28], it was emphasized that the superluminal propagation is a predeterminate result of a wave interference phenomenon where different frequency components of a pulse interference each other, and Pereyra and Simanjuntak [29] concluded unambiguously that the apparent superluminal effect is inherent to the electromagnetic theory. Therefore the superluminal group-velocity phenomenon is actually allowed by the Maxwell's equations. For the first viewpoint, if the signal velocity would equal the group velocity and be larger than $c$, we think, another important factor — noise — has to be considered. As emphasized by Wynne [14], any investigation into whether signal can travel superluminally or not *has to take into account noise*. For an information contained in a smooth light pulse, its signal velocity might be larger than $c$ or be negative without violating the causality, because *various random noise* (unpredictable), such as thermal noise at a temperature higher than absolute zero, and quantum noise existing at all situations, *is inevitable in the process of the pulse signal through any kind of medium* and *the noise prevents the possibility for any original signal back to the past without losing some parts of the signal*. That is to say, the receiver actually obtain the superluminal signal *at the price of losing some parts of the original signal*. In fact, it is known that, in normal dispersive media, the signal via a long-distance transmission always companies with noise, which can partially cover some parts of the original signal or totally obliterate the original signal [30]. ***This process is irreversible***. Some groups also argued that, for a smooth light pulse, the quantum noise may have limitation on superluminal pulse propagation [31-32], and Kuzmich et al. [10] further defined an operational signal velocity based on optical signal-to-noise ratio (SNR) and found that "the amplified quantum fluctuations introduce additional noise that effectively reduces the SNR in the detection of the signals carried by the light pulse". Clearly the noise plays an important role for keeping the causality. Therefore, for the first viewpoint, the superluminal signal velocity seems to be possible and compatible with the causality but at the price of *irreversibly losing* some of original signals due to noise.

Theoretcially, one has another choice, or say, the principle: the signal velocity in both normal and anomalous media exactly equals to $c$, independent of the transmission channels (media). In this case, the signal/information can be defined by an unpredictable or sudden change [1]. As said by Bigelow et al. [17], "since a true discontinuity would posses an infinitely broad frequency spectrum, it would propagate at velocity $c$, since no material could respond to arbitrarily large frequency components." Mathematically, one can create any perfect sudden change in wave packets, however in the strict sense it is practically impossible to produce an abrupt change since any abrupt change is always involved with an infinite-band spectrum. For the signal defined by any sudden change, an important but still indistinct problem is whether one has any way to modulate such a signal/information to provide an acceptable replica of it at the receiver in the communication systems. As show in Fig. 3(b), the sudden change is totally distorted after the first medium. In the practical applications one might generate a nonideal "sudden" change with a finite but very short transition time corresponding to a definite but very broad spectrum, i. e., the transition time $\Delta t$ is $0 < \Delta t \leq \varsigma$ (where $\varsigma$ is a finite but very small quantity), then its corresponding spectrum $\Delta\omega \propto 1/\Delta t$ has the range of $1/\varsigma \leq 1/\Delta t < \infty$. However, in order to transfer such a "sudden" change as a signal, its effective spectrum must be within the corresponding spectral response region of both the transmission channel (such as a fiber, a cable, and dispersive media) and the receiver, otherwise the signal distortion may inevitably appear due to the imperfect response of the transmission channel to the desired signal itself [30]. Actually, we think that, once the spectral bandwidth of the "sudden" change in experiments or communication systems is beyond the response region of the transmission channel, such "sudden" changes have the similar properties as that of the "noise", which cannot be predicted and with the spectrum beyond the response spectral width of the transmission channel. Therefore the speed of the "sudden" change naturally does not depend on the transmission channel. As pointed out by Bigelow et al. [17], the pulse edge (not a truly discontinuity) in their experiment should travel at velocity $c/n$ rather than at $c$, and Centini et al. [9] showed that the signal velocity (in terms of a threshold level) is always less than $c/n$, even though the peak propagates at the group velocity. If the refractive index would equal unity for these high-frequency components of the "sudden" change, the velocity $c/n$ would reduce to $c$ as observed by Stenner et al. [8, 16]. We think these experiments on the speed of the "sudden" change only tell us that the noise (including all unpredictable "sudden" changes) propagates at $c$ (or $c/n$).

In summary, we have discussed the possibilities of two kinds of pulse signals. For the first viewpoint, we think, it might be possible that the group velocity in fast-light cases could be the signal velocity at the price of losing some parts of original signals, therefore the transmission channel is very important to signal velocity. For the second viewpoint, we think, although there have several well-known experiments [8, 16-17] to verify the signal velocity of a discontinuity not larger than $c$, yet there is lack of experiment using the sudden change or discontinuity to encode any useful signal in any communication system. In real communication systems, a discontinuity in wave packets always leads to the strong distortion of the signal, which should be avoided during the propagation of the real signal. In fact, in order to have a high-quality signal transfer in transmission channel, the signal is always well modulated to match the bandwidths of both the transmission channel and the receiver [30]. For a signal having a well-defined shape, its propagating properties are always determined by the transmission channel (medium). The experiment done by Wang et al. [20] compelled one to think out of ways to discard the group velocity as the signal velocity in order to avoid the breaking of the causality. Whether is this right or not? Cautiously speaking, it is too far to have a final conclusion due to no directly experimental verification on the signal transfer by using the discontinuity or sudden change.

This work was supported by the National Nature Science Foundation of China (Nos. 10604047 and 10547138), by Zhejiang Province Scientific Research Foundation (G20630 and G80611).

**FIGURE CAPTIONS**

Fig. 1. Schematic of a pulse through an absorbing medium and a gain medium

Fig. 2. (Color online) The evolutions of (b)-(c) the relative intensity for a smooth light pulse through the absorbing and gain media. Solid and dashed lines denote the light pulses through the media and vacuum, respectively. Points A, B and C denote the peak times at different positions, respectively. The input pulse envelope is given by $E_{in}(0,t) = A\exp[-t^2/(2\tau^2)]$, where $A$ is a constant and $\tau = 1.2$ μs is the pulse half-width, and the pulse's carrier frequency is $\omega_0/2\pi = 3.5\times 10^{14}$ Hz. The media's parameters are as follows: $\Delta_a/2\pi = \Delta_g/2\pi = 1.35\times 10^6$ Hz, $\gamma_a/2\pi = \gamma_g/2\pi = 0.46\times 10^6$ Hz, and $M_a/2\pi = -M_g/2\pi = -2.262$ Hz. Note that Fig. 2 (a) shows the shape of the incident pulse and the solid curve in (b) is amplified by a factor 5.

Fig. 3. (Color online) The evolutions of (a)-(c) the relative intensity for a pulse with a sudden change through the absorbing and gain media. Solid and dashed lines denote the light pulses through the media and vacuum, respectively. Points A, B and C denote the peak times at different position, respectively. The input pulse envelope is given by $E_{in}(0,t) = A\exp[-t^2/(2\tau^2)]$ for $t \leq -\tau$, otherwise $E_{in}(0,t) = 2A\exp[-t^2/(2\tau^2)]$, where $A$ is a constant and $\tau = 1.2$ μs is the pulse half-width, and the pulse's carrier frequency is $\omega_0/2\pi = 3.5\times 10^{14}$ Hz. The media's parameters are as follows: $\Delta_a = \Delta_g = 0$, $\gamma_a/2\pi = \gamma_g/2\pi = 1.5\times 10^6$ Hz, and $M_a/2\pi = -M_g/2\pi = -2.262$ Hz. Note the sudden change $Q$ is located at time $t = -\tau$ of the input pulse, and the numerical integral range in the frequency domain is $[-120/\tau, 120/\tau]$, which has the sufficient accuracy for showing the sudden change. Note that Fig. 3 (a) shows the shape of the incident pulse and the solid curve in (b) is amplified by a factor 10.

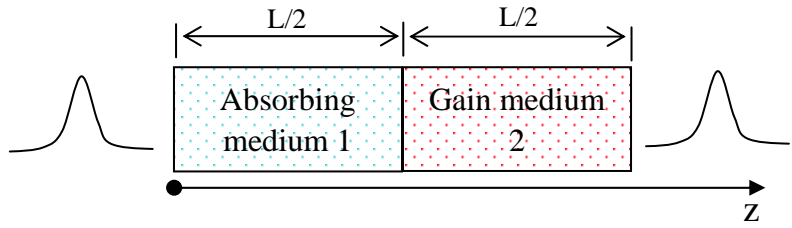

**FIG. 1**

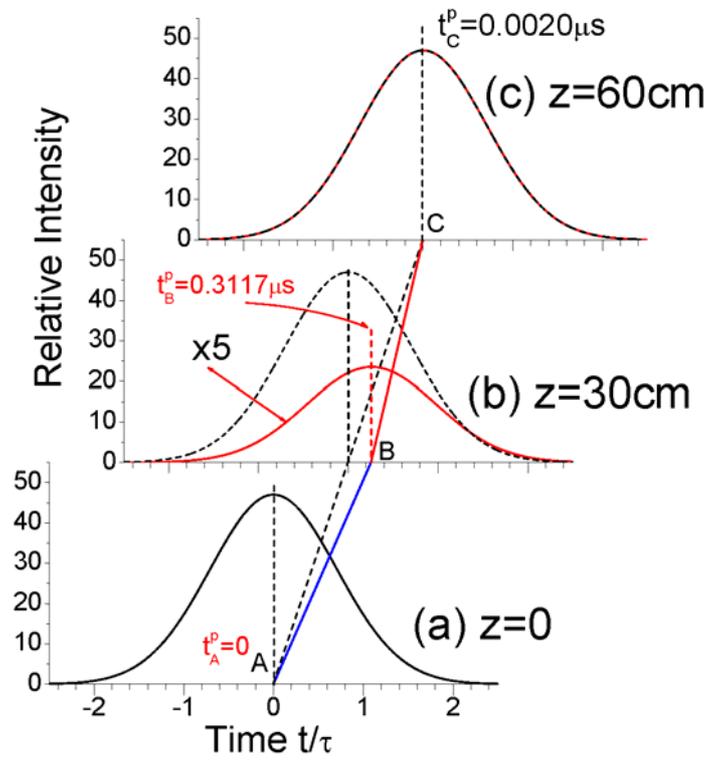

**FIG. 2**

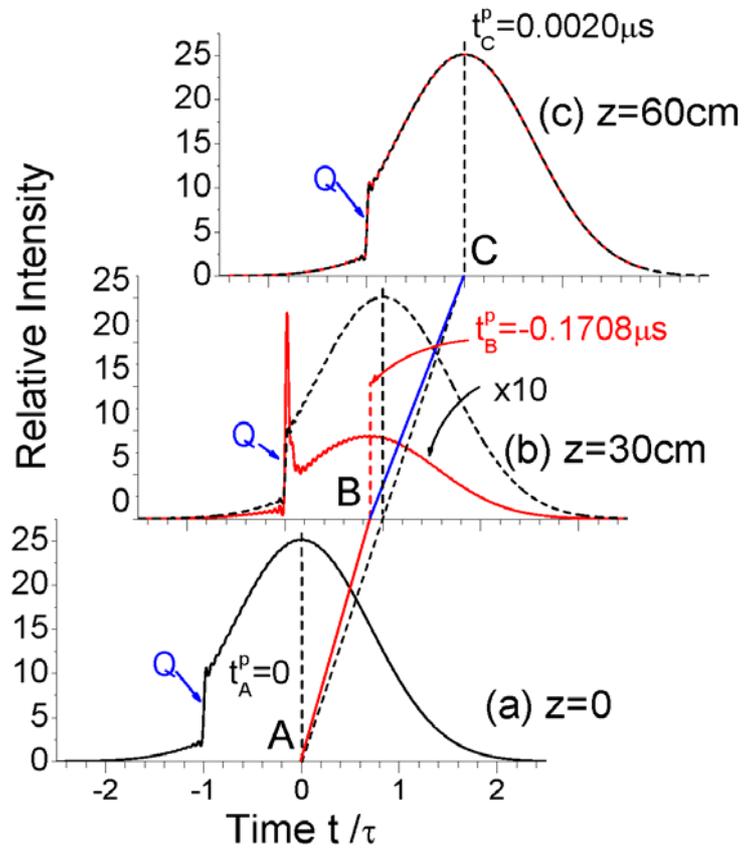

**FIG. 3**